\documentclass[%
 reprint,
superscriptaddress,
nofootinbib,
 amsmath,amssymb,
 aps,
]{revtex4-1}
\usepackage{xcolor}
\usepackage{graphicx}
\usepackage{dcolumn}
\usepackage{booktabs}
\usepackage{bm}

\usepackage{rotating} 

\usepackage{url}
\usepackage[linktocpage]{hyperref}

\hypersetup{
	colorlinks,
	linkcolor={red!60!black},
	citecolor={green!50!black},
	urlcolor={blue!80!black}
}

\usepackage{array} 

\usepackage{orcidlink}

\definecolor{amethyst}{rgb}{0.6, 0.4, 0.8}

\usepackage{soul}

\usepackage{bm}

\newcommand{\Mpc}{\, \text{Mpc}}
\newcommand{\kmsMpc}{\,{\rm km\,s^{-1}\,Mpc^{-1}}}

\newcommand{\ob}{\omega_\mathrm{b}}
\newcommand{\ocdm}{\omega_\mathrm{cdm}}

\newcommand{\Neff}{N_{\mathrm{eff}}}
\newcommand{\Nur}{N_{\mathrm{ur}}}
\newcommand{\sumnu}{\sum m_{\nu}}
\newcommand{\gnu}{g_\nu}

\begin{document}

\preprint{APS/123-QED}

\title[Resonant neutrino self-interactions]{Resonant neutrino self-interactions: insights from the full shape galaxy power spectrum}

 \author{{Hernán E. Noriega}\orcidlink{0000-0002-3397-3998}}\email{henoriega@icf.unam.mx}
    \affiliation{Instituto de Ciencias F\'isicas, Universidad Nacional Aut\'onoma de M\'exico, Apartado Postal 48-3, 62251 Cuernavaca, Morelos, M\'exico}
    \affiliation{Instituto de Física, Universidad Nacional Autónoma de México, Apartado Postal 20-364, 01000, D.F, México.}

    \author{{Josue De-Santiago}\orcidlink{0000-0002-1163-3730}}\email{Josue.desantiago@cinvestav.mx}
    \affiliation{ Departamento de F\'isica, Centro de Investigaci\'on y de Estudios Avanzados del Instituto Politécnico Nacional.  
                    Apartado Postal 14-740, 07000, Ciudad de M\'exico, M\'exico}
    \affiliation{Secretar\'ia de Ciencia, Humanidades, Tecnolog\'ia e Innovaci\'on,
    			Av. Insurgentes Sur 1582, 03940, Ciudad de M\'exico, M\'exico}

    \author{\mbox{Gabriela~Garcia-Arroyo}\orcidlink{0000-0002-0599-7036}}\email{arroyo@icf.unam.mx}
    \affiliation{Instituto de Ciencias F\'isicas, Universidad Nacional Aut\'onoma de M\'exico, Apartado Postal 48-3, 62251 Cuernavaca, Morelos, M\'exico}

   \author{{Jorge Venzor}\orcidlink{0000-0003-4510-2089}}\email{jorge.venzor@cinvestav.mx}
    \affiliation{ Departamento de F\'isica, Centro de Investigaci\'on y de Estudios Avanzados del Instituto Politécnico Nacional. 
                    Apartado Postal 14-740, 07000, Ciudad de M\'exico, M\'exico}
     \affiliation{Tecnologico de Monterrey, Escuela de Ingenier\'ia y Ciencias,
Av. Heroico Colegio Militar 4700, 31300, Chihuahua, Chihuahua, M\'exico}
    		   
\author{{Abdel P\'erez-Lorenzana}\orcidlink{0000-0001-9442-3538}}\email{abdel.perez@cinvestav.mx}
    \affiliation{ Departamento de F\'isica, Centro de Investigaci\'on y de Estudios Avanzados del Instituto Politécnico Nacional. 
                    Apartado Postal 14-740, 07000, Ciudad de M\'exico, M\'exico}


\date{\today}

\begin{abstract}
This paper investigates resonant neutrino self-interactions in cosmology by employing, for the first time, the effective field theory of large-scale structure to model their impact on the matter distribution up to mildly nonlinear scales. We explore a broad range of mediator masses in two main analyses: one combining BOSS Full Shape (FS) galaxy clustering with Big Bang Nucleosynthesis (BBN), and another combining FS with Planck Cosmic Microwave Background (CMB) data. 
Our results place the strongest cosmological constraints to date
on the resonant self interactions when using FS+Planck data,
reaching up to $g_\nu< 10.8 \times 10^{-14}$ at 95\% confidence for the 1 eV mediator. Notably, FS+BBN can constrain the interaction for the
10 eV mediator independently of CMB data, yielding $g_\nu < 7.33 \times 10^{-12}$ at 95\% confidence.
Our results suggest that resonant neutrino self-interactions are unlikely to resolve existing cosmological tensions within the standard $\Lambda$CDM framework.

\end{abstract}

\maketitle


\section{\label{sec:introduction} Introduction}

The ability of cosmology to act as a trustworthy neutrino probe has been under scrutiny over the last decades.
Thus far, it has worked well as an indirect indicator of neutrino physics, because the early Universe can test regions of the parameter space unreachable to neutrino experiments, from high-density environments at Big Bang Nucleosynthesis (BBN) to very low energies $\mathcal{O}$(keV-meV) at the cosmic microwave background (CMB) and large-scale structure formation \cite{Planck:2018vyg,adame2024desiBAO,adame2024desiFS}. Many of those phenomena occur over large periods, which makes the Universe behave as a long-exposure experiment; this ultimately leaves unique traces on cosmological observables, which works to constrain low-energy neutrino properties including their total mass \cite{Archidiacono_2020,GARIAZZO2023,Green:2024xbb,loverde2024massive,Noriega:2024lzo,Craig2024, Elbers:2024sha,Wang:2024hen,bertolez2024origin} and self-interactions. 

In recent years, cosmology has been able to test neutrino models with greater accuracy thanks to its ever-increasing data, new observables, and theoretical advances. These improvements have also helped break the degeneracy between neutrinos and other cosmological parameters. 
Modern studies can test neutrino physics beyond the standard model and, in particular, neutrino nonstandard interactions (NSI) with mediators with masses below $\sim \mathcal{O}$(MeV).
In these scenarios, there could be changes to the neutrino background \cite{Beacom2004,Hannestad2005,Chacko2020,Escudero2020relaxing,Esteban2021,Barenboim_2021,Abellan2022,Chen2022,Sandner2023,Venzor2021,Huang2018}, or its perturbations \cite{Bell2006,Archidiacono2014,CyrRacine2014,Oldengott_2015,Oldengott_2017,Lancaster2017,Kreisch2020,Kreisch:2022zxp,Park2019,Choudhury2021,Blinov2019,brinckmann2020self,mazumdar2022flavour,Huang2021,Das:2023npl,RoyChoudhury2022,Taule2022,Pal2024yom,Forastieri2015,Forastieri2019,Venzor2022,Venzor:2023aka,calabrese2025atacama,Barenboim2019inflation,Bostan_2024}; in particular, because the interaction reduces the neutrino free flow. 

Non free-streaming neutrinos have a scale-dependent impact that imposes a rich phenomenology to structure formation, which highly depends on the parameters related to the NSI.
Studies have obtained strong upper bounds on the interaction couplings because the reduced free-streaming requires
an extra radiation component to compensate for this effect (see for instance, \cite{Venzor2022}).
However, a long-standing anomaly has emerged for the \textit{heavy mediator} scenario, where the data prefer a nonnull interaction, this includes CMB temperature fluctuations and the full-shape (FS) galactic power spectrum (except for Planck polarization data) \cite{Kreisch2020,Kreisch:2022zxp,cerdeno2023constraints,he2023self,Camarena:2024daj,Poudou:2025qcx,He:2025jwp}.
Furthermore, the significance of the anomaly is high and can be tested with terrestrial neutrino experiments, even extending the model to include flavor physics \cite{Heurtier2017,Brune2019,Das2021,Schoneberg2019,Brdar2020,Esteban2021_probing,Cerdeno2021,Deppisch2020,Bustamante2020,Akita2022,Dutta:2023fdt,Medhi:2021wxj,Ge:2021lur,Escrihuela:2021mud,Escudero2020_CMB_search,Lyu2021,Seto2021,Fiorillo2022,Dutta2023,Denton:2024upc,bhupal2024new,Medhi:2023ebi,Zhang:2024meg,Berryman2018,DeRomeri:2024iaw,BLUM2018,foroughi2025enabling,Shalgar2021,Suliga2021,chang2022towards,Wang:2025qap,das2025impostor,akita2024limitsheavyneutralleptons,Fiorillo2024_small_impact,Fiorillo2024_theoretical}.

In Ref. \cite{Venzor:2023aka}, we first introduced the study of neutrino self-interactions in the resonant regime—where the interaction rate peaks at redshifts around $\sim 4\times 10^3 (m_{\phi}/{\rm eV})$—for linear cosmological perturbation data, where we tested mediator masses in the range $m_\phi = 10^{-2}$-$ 10^{2}$ eV. 
We used data from the Planck collaboration along with expansion history measurements, including baryon acoustic oscillation (BAO) data from the baryon acoustic oscillation spectroscopic survey (BOSS) and local measurements of $H_0$.
In that article, we found that resonant neutrino self-interactions (RNSI) moderately reduce the $H_0$ tension, and it mildly prefers a nonnull interaction, when using the Planck+BAO+$H_0$ dataset, if the mediator mass ranges between $0.5$ eV and $10$ eV.
Complementary constraints on neutrino resonant self-interactions arise from astrophysical observations, in particular from supernova neutrino signals, where interactions affect the duration of the neutrino burst, and from the diffuse supernova neutrino background (DSNB), which has been proposed as a probe for detecting such resonances \cite{Creque2021,chang2022towards,cerdeno2023constraints,Wang:2025qap,Fiorillo2024_small_impact,Fiorillo2024_theoretical}.
However, these resonances occur at energies much higher than the cosmological ones.

The neutrino self-interactions modify the shape of the matter power spectrum due to its scale-dependent effects. This allows for CMB-independent constraints on the interaction coupling, as previously done in \cite{Camarena2023} to constrain heavy mediators by combining FS with BBN data. Other studies have explored the constraints on heavy mediators \cite{he2023self,Camarena:2024daj,He:2025jwp,Poudou:2025qcx}, and model-independent NSI \cite{Pal2024yom,Kumar_2022}, by combining FS with CMB data.  
In this work, for the first time, we investigate the less-explored RNSI region, considering CMB-independent constraints and those obtained from FS combined with CMB data. To this end, we explore a wide range of mediator masses, covering a broad and phenomenologically rich region of parameter space.
Finally, we compare our results with previous constraints from BAO and CMB data.

The article is organized as follows. In Sec. \ref{sec:lperturbations} we present and review the equations of the RNSIs and analyze the impact of neutrino self-interactions on the matter power spectrum at both linear and mildly nonlinear scales.
Next, in Sec. \ref{sec:data} we present the datasets and outline the methodology.
Lastly, we discuss our results and give our concluding remarks in Secs. \ref{sec:results} and \ref{sec:conclusions}, respectively.

\section{\label{sec:lperturbations} Cosmological perturbations}

To track the dynamics of self-interacting neutrinos, it is necessary to solve the Boltzmann hierarchy with collision terms.
In the resonant region, the temporal evolution rate of the interaction $\Gamma$ depends on the mass of the (scalar) particle, $m_\phi$, that mediates the interaction, as calculated in \cite{Venzor:2023aka} following the formalism of Refs. \cite{GONDOLO1991,Vassh2015}.
Here, it is important to recall that other inelastic processes are being ignored and only the neutrino self-interactions are considered.

In the conformal Newtonian gauge \cite{Ma1995}, the linear-order hierarchy for the massive neutrino perturbations $\Psi$ is
\cite{Hannestad2000,Archidiacono2014,Forastieri2015, Venzor:2023aka}:
\begin{eqnarray} 
\dot{\Psi}_0 &=& -\frac{qk}{\epsilon}\Psi_1- \dot{\phi}\frac{d \ln f_0}{d \ln q} \, , \label{eq:Boltzmann_l0}\\
\dot{\Psi}_1 &=& \frac{qk}{3\epsilon}(\Psi_0 -2\Psi_2) -\frac{\epsilon k }{3q}\psi \frac{d \ln f_0}{d \ln q}\, ,\label{eq:Boltzmann_l1}\\
\dot{\Psi}_{l \geq 2} &=& \frac{q k}{(2l+1)\epsilon}\left[l\Psi_{l-1}-(l+1)\Psi_{l+1}\right] -a \Gamma \Psi_l, \label{eq:Boltzmann_l2}
\end{eqnarray}
where the dots denote conformal time derivatives, $\phi$ and  $\psi$ are the metric scalar potentials, $f_0$ is the unperturbed Fermi-Dirac distribution, $\epsilon = \sqrt{q^2 +a^2m_{\nu}^2}$ is the comoving energy, and $q$ is the magnitude of the comoving momentum. The self-interaction conserves both energy and momentum but affects significantly the multipoles $l \geq 2$, this is of particular interest in cosmology given that this moment is directly associated with anisotropic stress,  which, through Einstein's equations, modifies the evolution of gravitational potentials and can leave a footprint in the CMB and the matter power spectrum (MPS), among other observables.

\begin{figure}
    \centering
    \includegraphics[width=.48\textwidth]{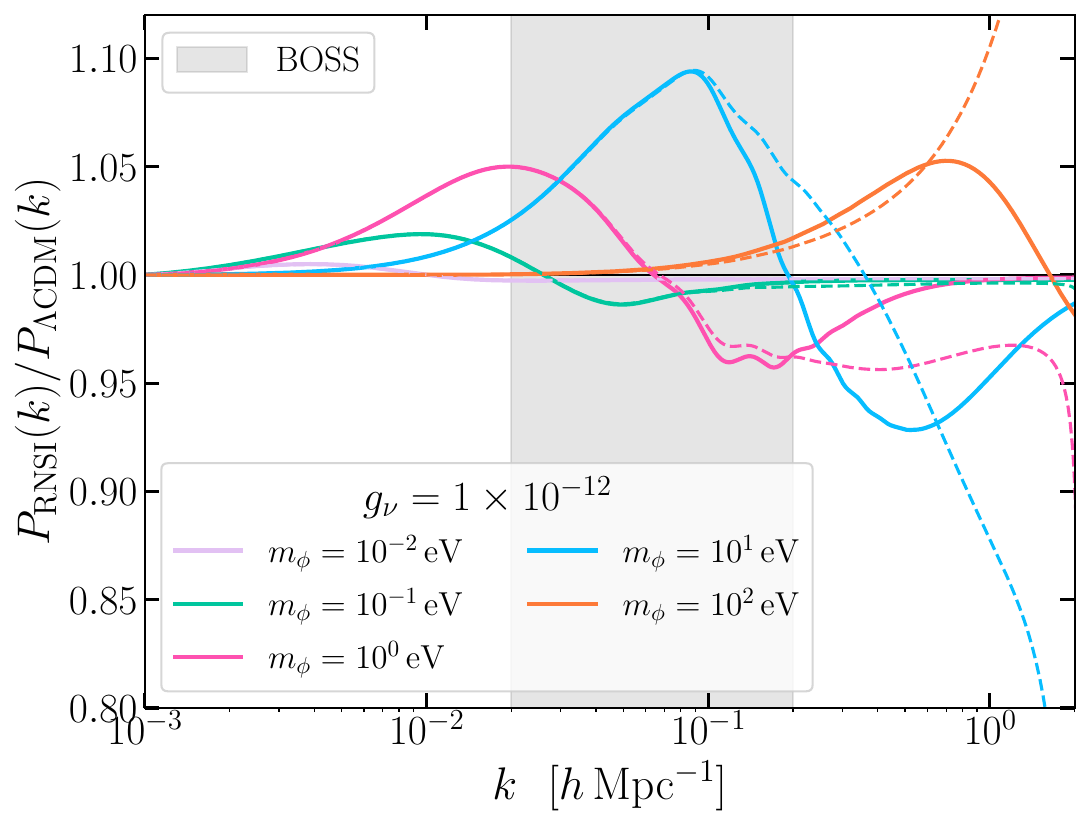}
    \caption{Ratio of the linear and nonlinear matter power spectra of RNSI with respect to the $\Lambda$CDM model at $z=0.38$. Each mediator mass $m_{\phi}$ is represented by a different color; the solid line corresponds to the linear comparison, while the dashed line shows the ratio between the real matter power spectra including contributions up to one-loop order. The shaded gray region highlights the scales probed by BOSS. 
    Through this work we adopt $k = 0.20\ h \Mpc^{-1}$ as the maximum wavenumber where one-loop perturbation theory is expected to remain reliable. Deviations beyond this scale are not physically meaningful and are shown for illustration only.
}
    \label{fig:ratio_mps}
\end{figure}

The RNSI introduces two parameters, $m_\phi$, and $g_\nu$, codified into $\Gamma$, given by
\begin{equation}\label{eq:scatt_rate}
    \Gamma= \frac{g_{\nu}^2 m_{\phi}^2}{24 \pi \zeta(3) T_\nu}F(m_{\phi}^2;T_\nu).
\end{equation}
Eq. \eqref{eq:scatt_rate} was computed in Ref. \cite{Venzor:2023aka} using the  M\o ller velocity approach with $\Gamma=\left< \sigma v_{\rm MOL}\right> n_{\nu}$. 
Here, the cross section of the self-interaction behaves as a Dirac delta function $\sigma \rightarrow (4\pi g_{\nu}^2 E_\nu^2)/({m_{\phi}^2}) \ \delta (4E_\nu^2-m_{\phi}^2)$ given the narrow width limit (which fits perfectly for the small $\gnu$ values relevant to this analysis).
In addition to $F(m_{\phi}^2;T_\nu)$, which is a well-behaved numerical function, notice that the resonant interaction rates scale as $\gnu^2$ instead of the non-resonant counterparts employed in other mediator masses approximations that scale as $\gnu^4$.
Finally, we recall that $\Gamma$ peaks around a redshift given by $\sim 4 \times 10^3 (m_{\phi}/{\rm eV})$, but dies down rapidly when the redshift is far from the $z_{\rm peak}$.

These two parameters ($m_\phi$, and $g_\nu$) modulate neutrino free-streaming and self-interaction coupling, leaving distinct signatures on linear and nonlinear scales depending on their combination (see Fig.~\ref{fig:ratio_mps} for a fixed $g_\nu$).
To account for these effects, we modify the Einstein-Boltzmann solver \textsc{CLASS} \cite{class2,class4} by incorporating Eqs.~(\ref{eq:Boltzmann_l0})--(\ref{eq:Boltzmann_l2}). We then merge it with \textsc{CLASS-PT} \cite{Chudaykin:2020aoj},\footnote{Recently, Ref.~\cite{Maus:2024sbb} (and references therein) demonstrated that different PT/EFT models and the codes implementing them, such as \textsc{CLASS-PT}, \textsc{FOLPS}, \textsc{PyBird}, \textsc{Velocileptors}, and \textsc{CLASS-OneLoop}, yield consistent results under equivalent settings.} an effective field theory (EFT)-based code that relies on Eulerian Perturbation Theory (Eulerian PT) and Einstein–de Sitter (EdS) kernels to compute the one-loop redshift-space galaxy power spectrum and its multipoles.

In principle, in the context of the RNSI model, the PT/EFT description should be modified. However, for a wide range of mediator masses, specifically $m_\phi \sim 1$-$10^2$ eV, the free-streaming is reached again before the matter-dominated era (see Fig.~1 from \cite{Venzor:2023aka}), as in the standard non-self-interacting neutrino case for realistic neutrino masses. Consequently, the changes induced in the gravitational field at horizon crossing only affect the initial shape of the linear power spectrum, without introducing significant contributions to the evolution of perturbations in the nonlinear regime \cite{he2023self, Camarena2023}. Moreover, the rate of neutrino self-interactions becomes efficient at high redshifts ($z \gtrsim 4000$) for the specified range of mediator masses, where nonlinear effects are negligible. Therefore, the standard PT/EFT implementation is well-suited for describing the late-time evolution of large-scale structures in the presence of self-interacting neutrinos.

Figure \ref{fig:ratio_mps} illustrates the deviations in the MPS between the RNSI and $\Lambda$CDM models for the explored mediator masses. We adopt the Planck best-fit cosmology \cite{Planck:2018vyg}, assuming a total neutrino mass of $0.06\, \text{eV}$. In the figure, solid and dashed lines represent the ratios of the linear and total (linear + nonlinear corrections) MPS, respectively, while the shaded gray region highlights the scales relevant for BOSS. The figure confirms that the characteristic bump observed in the MPS ratio progressively shifts to smaller scales as the mediator mass increases. 

Interestingly, for a fixed coupling $g_{\nu} = 10^{-12}$, the mediator mass $m_{\phi} = 10\, \text{eV}$ exhibits the largest deviation, with a peak at $k \sim 0.1\, h\, \text{Mpc}^{-1}$—a scale relevant for BOSS, where nonlinear effects start to become important.  At these scales, neutrino self-interactions suppress anisotropic stress, enhancing matter perturbations and producing a peak in the MPS. Beyond this peak, as $k$ increases, neutrinos gradually regain their free-streaming behavior, reducing their clustering contribution and leading to a progressive suppression of the MPS at small scales.

\section{\label{sec:data} Data and Methodology}
In this section, we briefly describe the datasets and their combinations, as well as the analysis setup employed throughout this work.

\subsection{Datasets}

\begin{enumerate}

    \item \textbf{BOSS:} 
    Anisotropic galaxy clustering drawn from the publicly available BOSS Data Release 12 (DR12) galaxy sample \cite{Dawson_2012, BOSS:2016wmc}, which contains redshifts and positions for 1,198,006 galaxies spanning the redshift range $0.2 < z < 0.75$.  We concentrate on the two non-overlapping redshift bins $0.2 < z_1 < 0.5$ and $0.5 < z_3 < 0.75$ with effective redshifts $z_{\rm eff} = 0.38$ and $z_{\rm eff} = 0.61$, respectively. Each redshift bin is further divided into two Galactic Caps, the North (NGC) and the South (SGC), covering a total sky area of 9,329 deg$^2$. To account for the anisotropy introduced by the assumed fiducial cosmology with $\Omega^{\rm fid}_m = 0.31$, we apply the Alcock-Paczyński effect \cite{Alcock:1979mp}.
    We used the products provided in \cite{Ivanov:2019pdj}.\footnote{\url{https://github.com/oliverphilcox/full_shape_likelihoods}} These include window-free galaxy power spectrum multipoles, as well as the covariance matrices extracted from $2\times2048$ realizations of the MultiDark-Patchy mock catalogs for the NGC and SGC \cite{Kitaura:2015uqa}. 
    The use of this dataset is simply referred to as FS.

    \item \textbf{BBN:} Complementary to BOSS data, we use measurements of the primordial abundances of helium and deuterium to constrain the baryon density parameter, $\omega_b$, and the effective number of relativistic species, $N_{\text{eff}}$.
    Given that BBN is not sensitive to perturbations and that the resonant self-interacting neutrino model becomes relevant after this epoch, it does not affect the nucleosynthesis process or introduce degeneracies with \( N_{\text{eff}} \).  For this dataset, the implementation we follow is the so-called PArthENoPE-standard proposed in \cite{Schoneberg:2019wmt}, which makes use of the deuterium measurements reported in \cite{Cooke:2017cwo} and helium data compilation from \cite{Aver:2015iza}.

    \item  \textbf{Planck:}
    This dataset includes the combined likelihoods of the temperature (TT), polarization (EE), and cross-correlation (TE) power spectra, along with CMB lensing. Specifically, we use the \texttt{Commander} and \texttt{SimAll} likelihoods for the low-$\ell$ temperature and polarization spectra (TTEE) and the \texttt{Plik} likelihood for the high-$\ell$ auto and cross spectra (TTTEEE), along with the \texttt{smica} lensing likelihood, all from Ref.~\cite{Planck:2019nip}. Throughout this paper, we refer to this combination simply as Planck.
\end{enumerate}
We investigate two different dataset combinations: FS+BBN and FS+Planck. Furthermore, we compare the latter with the constraints obtained in \cite{Venzor:2023aka} for the BAO+Planck combination.

\subsection{Analysis setup}

To test the RNSI model, we employ a modified version of the \textsc{CLASS} code \cite{class2,class4}, which feeds into the \textsc{CLASS-PT} code \cite{Chudaykin:2020aoj} to compute the first two non-zero multipoles of the galaxy power spectrum—the monopole and quadrupole—within the conservative range $0.01 < k / [h \, {\rm Mpc}^{-1}] < 0.20$. In this range, the PT/EFT models have been shown to yield unbiased results for mocks with BOSS-like volume and even larger volumes (see, e.g., \cite{Nishimichi:2020tvu, Noriega:2022nhf, Maus:2024sbb}). To explore the parameter space, we use the publicly available \textsc{MontePython} code \cite{Brinckmann:2018cvx, Audren:2012wb}, ensuring a Gelman-Rubin convergence criterion \cite{Gelman:1992zz} of $R - 1 \lesssim 10^{-2}$ across the Markov Chain Monte Carlo samples. Contour plots, posterior distributions, and summary statistics are generated using the \texttt{GetDist} package \cite{Lewis:2019xzd}.

\begin{table}[htb]
\begin{ruledtabular}
\begingroup
\renewcommand{\arraystretch}{1.25}
\begin{tabular}{lc}
Parameter  &  Prior \\
\hline
\textbf{Cosmological (FS+BBN)} \\
$\gnu$ & $\mathcal{U}(0.0, 10^{-6})$ \\
$\Nur$ & $\mathcal{U}(-3.0, 3.0)$ \\
$\sumnu\; [{\rm eV}]$ & $\mathcal{U}(0.0, 3.0)$ \\
$H_{0} \; [\kmsMpc]$ & $\mathcal{U}(50.0, 90.0)$ \\
$\ocdm$ & $\mathcal{U}(0.10 , 0.30)$ \\
$\ob$ & $\mathcal{U}(0.018, 0.030)$ \\
$\ln(10^{10} A_{s})$ & $\mathcal{U}(2.0, 4.0)$ \\    
$n_{s}$ & $\mathcal{U}(0.4, 1.5)$ \\

\multicolumn{2}{c}{\dotfill} \\

\textbf{Cosmological (FS+Planck)} \\
$100 \theta_s$ & $\mathcal{U}(-\infty , +\infty )$ \\
$\tau_{\rm reio}$ & $\mathcal{U}(0.004, +\infty )$ \\

\hline

\textbf{Nuisances} \\
$b_1 $ & $\mathcal{U}(1.0, 4.0)$ \\    
$b_2 $ & $\mathcal{U}(-10.0, 10.0)$ \\    
$b_{\mathcal{G}_2}$ & $\mathcal{U}(-10.0, 10.0)$ \\ 
$c^2_{0}$ & marginalized \\
$c^2_{2}$ & marginalized \\
$\tilde{c}$ & marginalized \\
$P_{\rm shot}$ & marginalized \\
\end{tabular}
\endgroup
\end{ruledtabular}
\caption{List of parameters and priors for the analysis. We use $\mathcal{U}$(min, max) to denote a uniform prior over the given range. The cosmological parameters at the top are those associated with the FS+BBN analysis, while those in the second block correspond to additional parameters introduced when combining FS with Planck data. This combined analysis samples $100 \theta_s$ instead of $H_{0}$. The bias parameters vary independently for each redshift bin and galaxy cap, while the counterterms and stochastic parameters are analytically marginalized over, as they enter at linear order in the power spectrum \cite{Philcox:2020zyp}. We compute $\Neff$, $\Omega_m$, and $\sigma_8$ as derived parameters.
}
\label{table:priors}
\end{table}

When fitting the FS+BBN dataset, we sample over the set of cosmological parameters $\{ H_0, \omega_{\rm cdm}, \omega_{\rm b}, A_s, n_s, \sum m_\nu, N_{\rm ur}, g_\nu \}$, which include the Hubble constant $H_0$, the present-day physical density of cold dark matter $\omega_{\rm cdm} = \Omega_{\rm cdm} h^2$ and baryons $ \omega_{\rm b} = \Omega_{\rm b} h^2$, the amplitude of primordial fluctuations $A_s$, the spectral index $n_s$, the total neutrino mass $\sumnu$, and the number of ultra-relativistic species $ N_{\rm ur}$. These define the $\Lambda$CDM model with $\sumnu$ and $N_{\rm ur}$ as free parameters. For brevity, we will refer to this as the $\Lambda$CDM model from here on, unless stated otherwise. We additionally include the parameter $g_\nu$ to account for neutrino self-interactions in the RNSI model. For fits that include CMB data, we further sample over the optical depth to reionization $ \tau_{\rm reio}$, and the acoustic angular scale $100 \theta_s$.

In the FS+BBN analysis, we explore a wide range of mediator masses, spanning $m_\phi \sim 1$-$10^2$ eV, for which the use of standard PT/EFT is justified; see Sec.~\ref{sec:lperturbations}. However, for the FS+CMB case, we extend the mass range to $10^{-2}$-$10^{-1}$~eV, while still employing EdS kernels. Although, in principle, this is not a fully valid approximation, we expect the dominant contribution to arise from CMB data, with the FS information being subdominant. In Sec.~\ref{sec:results}, we assess the consistency of this approach by comparing it to the results obtained from the BAO+Planck analysis~\cite{Venzor:2023aka}.

In addition to the cosmological parameters, we further include a set of nuisance parameters for each redshift bin and galaxy cap. These comprise the galaxy biases $b_1, b_2$ and $b_{\mathcal{G}_2}$, as well as the EFT and stochastic parameters $c^2_0, c^2_2, \tilde{c}$ and $P_{\rm shot}$. The cubic bias is fixed to $b_{\Gamma_3} = \tfrac{23}{42}(b_1 - 1)$, following its coevolution prediction \cite{Desjacques:2016bnm}. We analytically marginalize over the EFT and stochastic parameters, as they enter at linear order in the theory \cite{Philcox:2020zyp}. The setups and priors for our two main analyses, FS+BBN and FS+Planck, are summarized in Table~\ref{table:priors}.

\begin{figure*}
    \centering
    \includegraphics[width=1\textwidth]{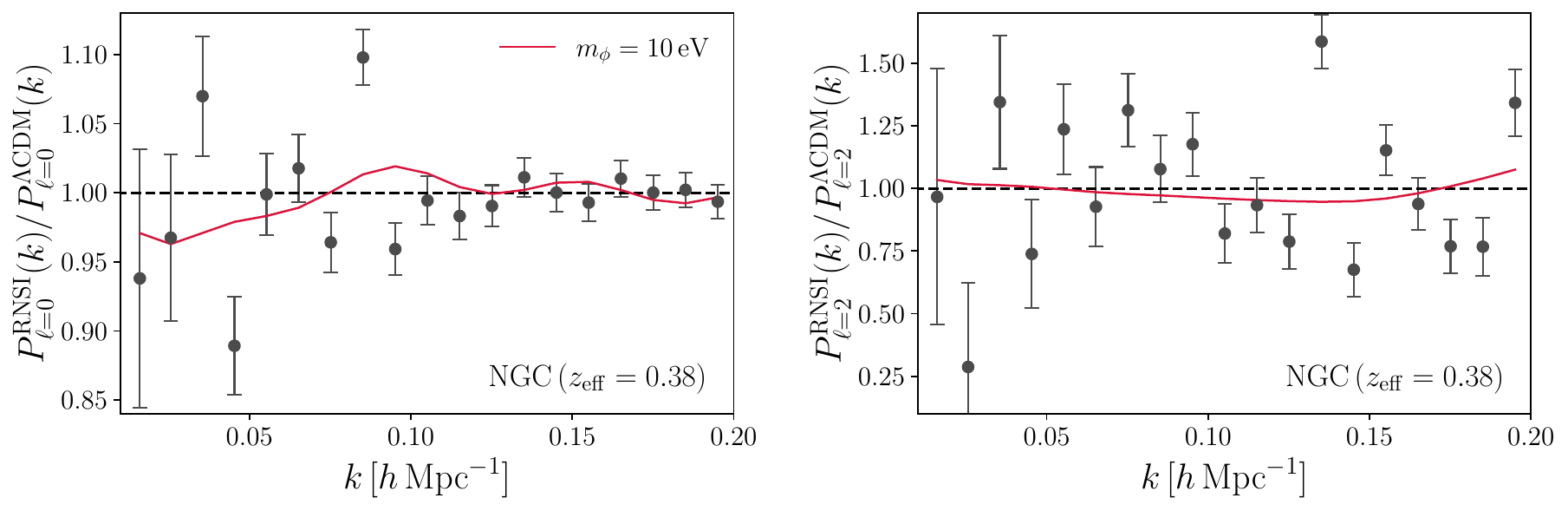}
\caption{Best fits for the redshift-space galaxy power spectrum monopole (left) and quadrupole (right) on the FS+BBN dataset. We present the RSNI best fit for $m_\phi = 10\,$ eV and compare it to $\Lambda$CDM. 
The data shown correspond to the NGC at $z_{\rm eff} = 0.38$. 
}
    \label{fig:betsfit}
\end{figure*}

\section{\label{sec:results}Results}

We assume three degenerate neutrino mass states, consistent with the latest Planck results \cite{Planck:2018vyg}. Constraints on the cosmological parameters of interest are summarized in Table~\ref{tab:constraints}, where the uncertainties are quoted at the $68\%$ confidence level (c.l.), and upper limits at the $95\%$ c.l.
The top section shows the results for the FS+BBN dataset combination, considering mediator masses of $1, 10$ and $100~\rm{eV}$, while the bottom section presents those for FS+Planck, which includes mediator masses of $10^{-2}$ and $10^{-1}$~eV. Although the perturbative approach may not be fully valid for such small mediator masses, we expect the dominant contributions to arise from the CMB.

Focusing first in the FS+BBN results, a striking difference arises in the constraint on the constant coupling, $g_{\nu}$, for the mediator masses  $m_{\phi}$ we explored.
Specifically, only for $m_{\phi} = 10\,\mathrm{eV}$ is the coupling parameter constrained, with $g_\nu < 7.33 \times 10^{-12}$ (95\% c.l.).
For other mediator masses, the Markov chains saturate the prior, and fail to constrain the coupling.
These results support our findings in Fig.~\ref{fig:ratio_mps}, which show that $m_{\phi}= 10\,\rm{eV}$ induces a more pronounced deviation in the total MPS at scales probed by BOSS than other mediator masses. This larger deviation enables meaningful constraints on $g_\nu$, while smaller deviations for other mediator masses lead to prior saturation and unconstrained $g_\nu$.

The best-fit RNSI model with $m_{\phi}=10\, \rm{eV}$ offers a better fit to the redshift-space galaxy power spectrum multipoles, particularly by more accurately reproducing their oscillatory features—most notably in the monopole. This is illustrated in Fig. \ref{fig:betsfit}, which shows the ratios of the redshift-space galaxy power spectrum multipoles of the best-fit RNSI model with $m_{\phi}=10\, \rm{eV}$ to those of $\Lambda$CDM, at $z_{\rm{eff}}=0.38$, for the BOSS NGC data. The left and right panels correspond to the monopole and quadrupole, respectively.

\begin{figure*}
    \centering
    \resizebox{\linewidth}{!}{
        \includegraphics[height=0.37\linewidth]{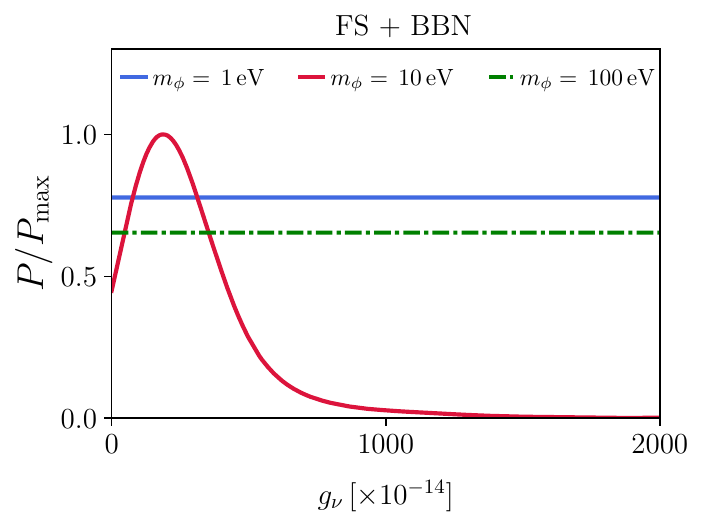}
        \includegraphics[height=0.37\linewidth]{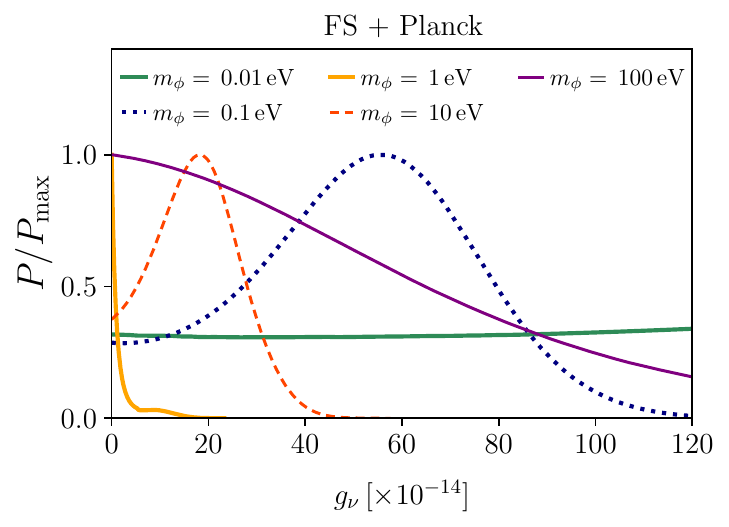}
    }
    \caption{one-dimensional marginalized posterior distributions of $g_\nu$ for different mediator masses in the FS+BBN (left) and FS+Planck (right) analyses. In the FS+BBN case, the mediator masses $m_\phi = 1, 100\, {\rm eV}$ completely saturate their priors, preventing the detection of neutrino self-interaction signature. Consequently, these values are not reported in the companion Table~\ref{tab:constraints}, where they are instead indicated by a ‘---’ symbol. In contrast, for FS+Planck,  $g_\nu$ is detected for all mediator masses. However, for visualization purposes, we truncate the scale. The corresponding triangle plots are shown in Fig.~\ref{fig:triangulars}.
    } 
    \label{fig:1D-gnu}
\end{figure*}

 The strength of this constraint is further illustrated by the one-dimensional marginalized posteriors of $g_\nu$ shown in Fig.~\ref{fig:1D-gnu}, for all considered mediator masses, in the FS+BBN (left panel) and FS+Planck (right panel) analyses.
In the FS+BBN case, the posterior for $m_{\phi}=10~\rm{eV}$ is sharply peaked, while those for  $m_{\phi}=1$ and $100~\rm{eV}$ saturate the prior range, indicating that the scales tested by BOSS data generally do not overlap with those where RNSI interactions induce significant deviations, and consequently there is a lack of constraining power for these particular mediator masses; accordingly, in Table \ref{tab:constraints} we do not report any limit on $g_{\nu}$ for  $m_{\phi}=1$ and $100~\rm{eV}$.

\begin{figure}
    \centering
    \includegraphics[width=1\linewidth]{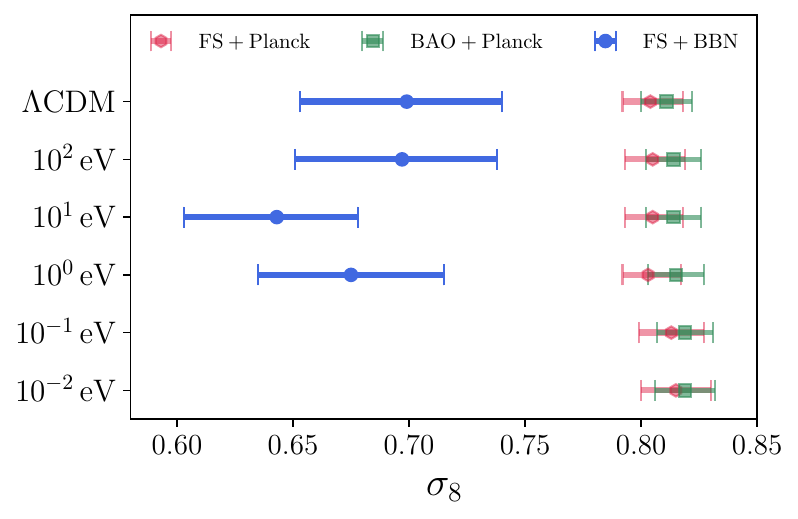}
    \caption{Constraints on $\sigma_8$ derived from the combination of different datasets and mediator masses. The error bars indicate the 68\% confidence level, while the markers around the middle represent the mean values. BAO+Planck constraints come from Ref. \cite{Venzor:2023aka}.}
    \label{fig:sigma8}
\end{figure}

Beyond the coupling constraints, the remaining cosmological parameters are well determined, see Fig. \ref{fig:triangulars}. For instance, the results reflect that the inferred values of the amplitude of matter fluctuations at $8 h\, {\rm Mpc^{-1}}$ ($\sigma_8$) are lower for the RNSI models compared to the $\Lambda$CDM scenario by up to 8\%, when using FS+BBN. While the estimations for this dataset are consistently lower—regardless of the specific model or mediator mass—than those from the FS+Planck and Planck+BAO (see Fig. \ref{fig:sigma8}).
These results reflect the trend of low-redshift data to prefer lower values of $\sigma_8$, which is compatible with a larger sum of neutrino masses, as indicated by the constraints presented in Table~\ref{tab:constraints}.

Moreover, within the FS+BBN results it is noticed that the case  with $m_{\phi}=10~\rm{eV}$ yields the lowest $\sigma_{8}$ value. This can be understood from the fact that $\sigma_8$ is most sensitive to the matter power spectrum around $k \sim 0.1\,\mathrm{Mpc}^{-1}$, where the suppression induced by RNSI is strongest for this particular mass, in line with the tightest constraint on $g_\nu$ found for this case. We note, however, that trends in $\sigma_8$ should be interpreted with caution, as their apparent tension can be misleading depending on dataset combinations and modeling assumptions~\cite{forconi2025illustrating,Sanchez2020}.
Furthermore, FS+BBN provides an alternative determination of $H_0$ that is independent of both CMB and supernova data. The constraints obtained from FS+BBN accommodate slightly higher values than those inferred by the Planck Collaboration~\cite{Planck:2018vyg} (see Table~\ref{tab:constraints}), while remaining statistically consistent with SH0ES. The inferred $H_0$ values in the RNSI scenarios are overall similar to those obtained for $\Lambda$CDM.

We now turn to the FS+Planck results. As shown in the right panel of Fig.~\ref{fig:1D-gnu} and the bottom section of Table~\ref{tab:constraints}, the inclusion of CMB data significantly tightens the constraints on $g_\nu$, now providing meaningful constraints for every mediator mass considered—whereas the FS+BBN combination was only able to constrain $g_\nu$ for $m_\phi = 10~\mathrm{eV}$.
%
 In this analysis,  the strongest constraint on $g_\nu$ is obtained for $m_{\phi}=1\,\text{eV}$, in agreement with the previous findings \cite{Venzor:2023aka} based on Planck+BAO data. 
 
\begin{figure*}
    \centering
    \resizebox{\linewidth}{!}{
        \includegraphics[height=0.37\linewidth]{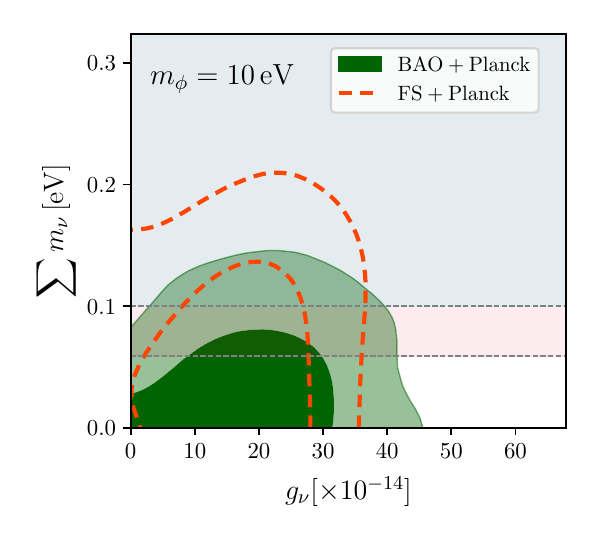}
        \raisebox{4mm}{\includegraphics[height=0.328\linewidth]{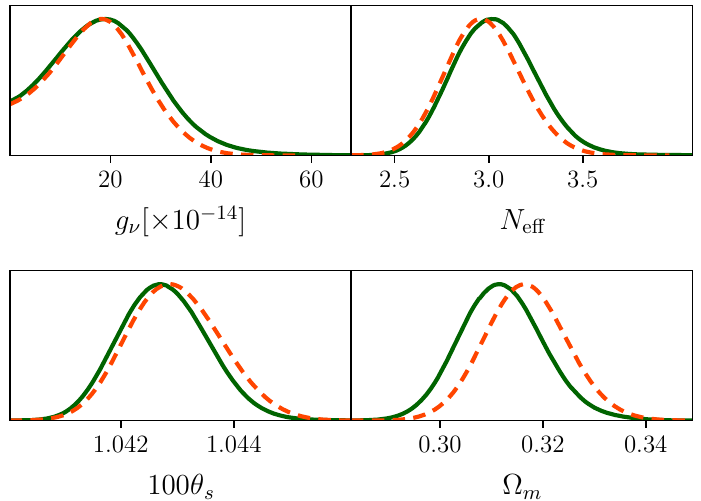}}
    }
    \caption{(left) Comparison of the two-dimensional constraints in the $\sum m_\nu$-$g_\nu$ parameter space from BAO+Planck and FS+Planck. The horizontal dashed lines and shaded regions indicate the minimal masses for the normal hierarchy ($\sum m_\nu > 0.059\, {\rm eV}$) and inverted hierarchy ($\sum m_\nu > 0.10\, {\rm eV}$). (right) Comparison of the one-dimensional marginalized posterior distributions for other cosmological parameters. The results are shown for a mediator with $m_\phi = 10\, {\rm eV}$, though similar trends are observed for other mediator masses.
    }
    \label{fig:2D_gnu-sumnu}
\end{figure*}

\begin{figure*}
    \centering
    \resizebox{\linewidth}{!}{
        \includegraphics[height=0.45\linewidth]{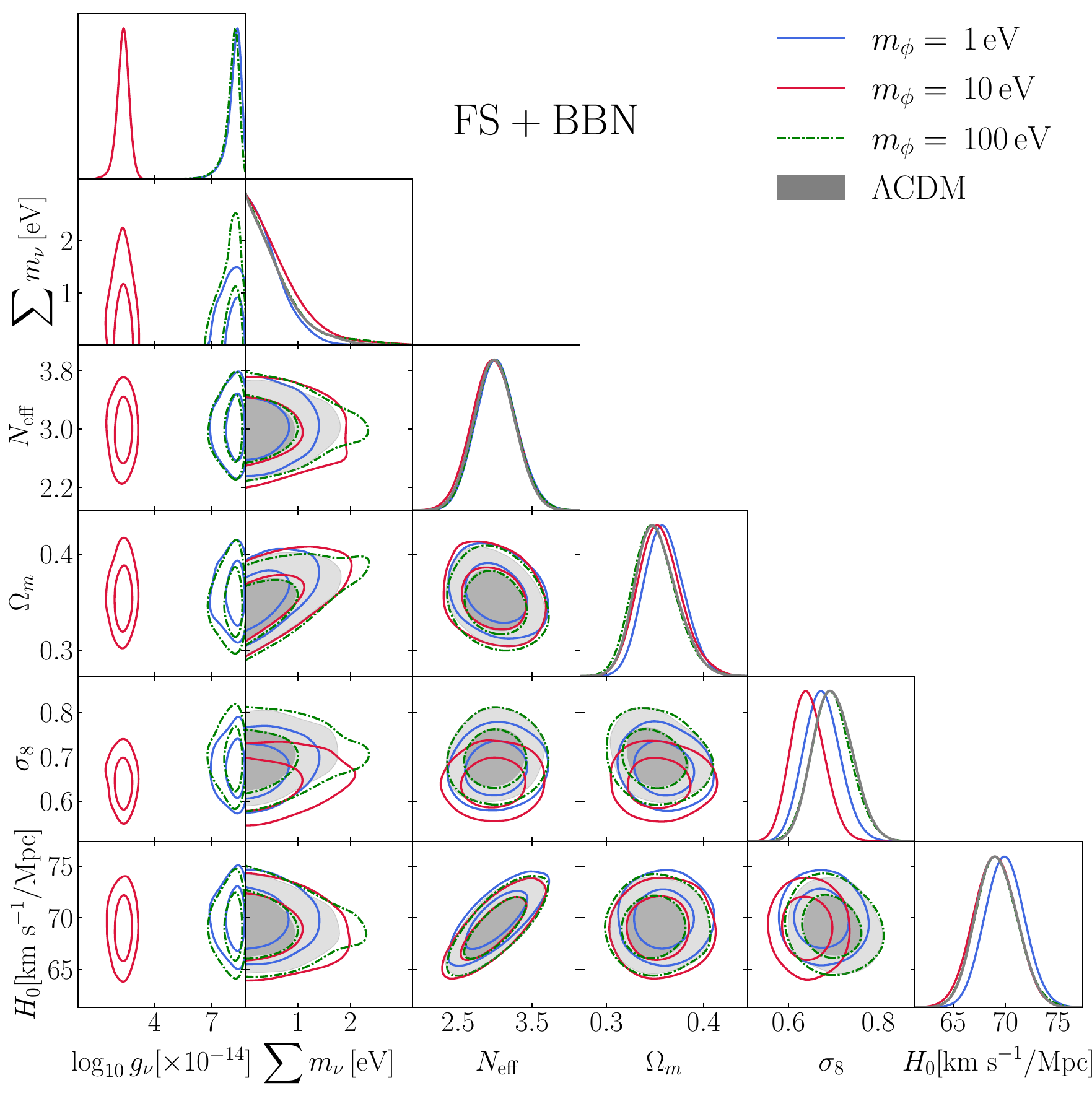}
        \includegraphics[height=0.45\linewidth]{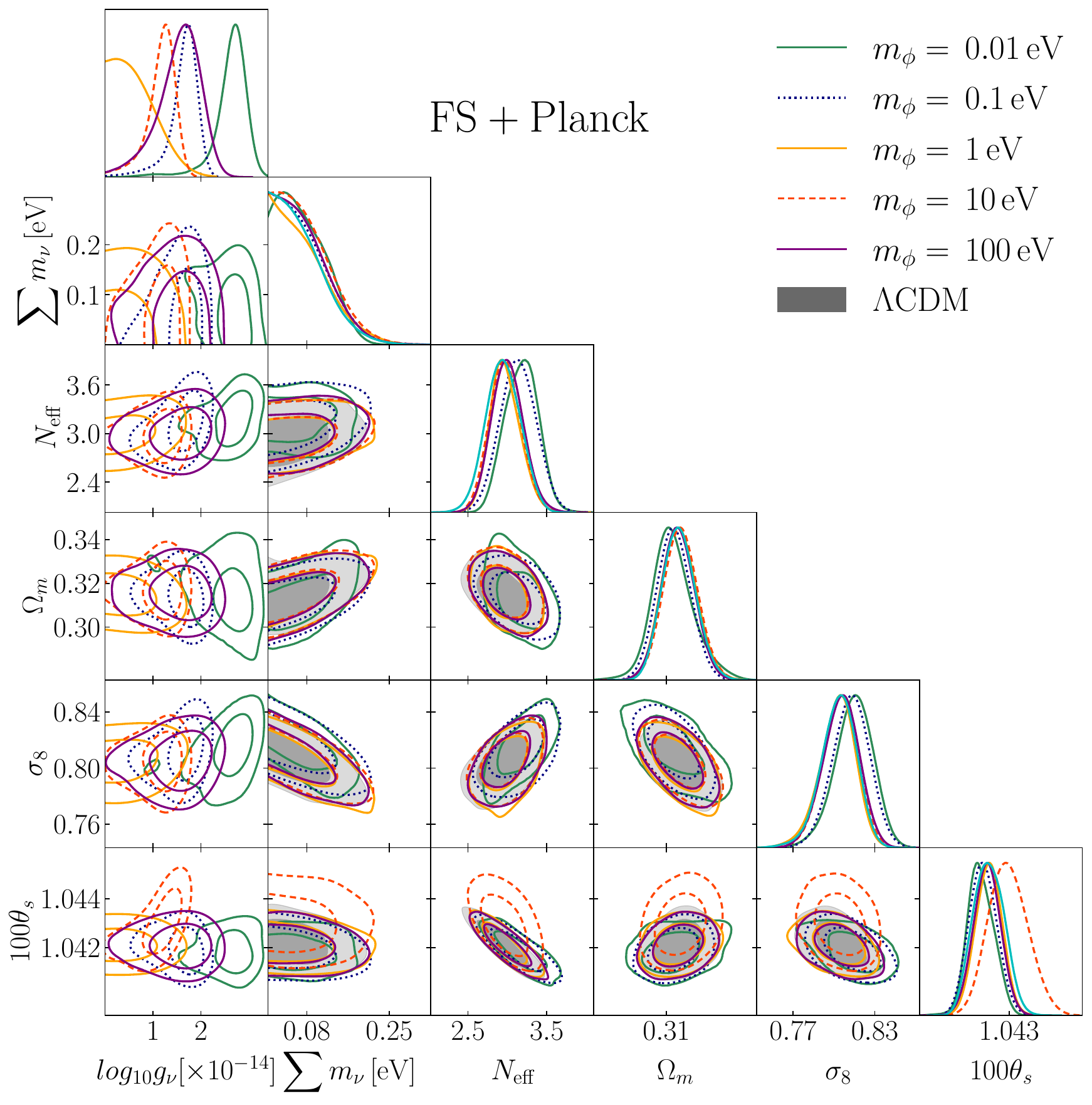}
    }
    \caption{Posterior distributions at the 68\% and 95\% confidence levels for different mediator masses in the FS+BBN (left) and FS+Planck (right) analyses. For comparison, we have also included the posterior distributions of the $\Lambda$CDM model with free $\sumnu$ and $N_{\rm eff}$ parameters. Note that the posterior of $g_\nu$ is presented on a logarithmic scale. As mentioned previously in Fig.~\ref{fig:1D-gnu}, $m_\phi = 1, 100\, {\rm eV}$ saturates the priors on $g_\nu$ for the FS+BBN case. The corresponding cosmological parameter constraints are presented in Table~\ref{tab:constraints}.
    } 
    \label{fig:triangulars}
\end{figure*}

Nevertheless, it is important to note that the constraint for $m_{\phi}=10\,\text{eV}$ in the FS+BBN case remains competitive with that obtained from FS+Planck, suggesting that FS data already provided meaningful bounds at this mass scale. Furthermore, the upper limit on $g_{\nu}$ is tighter in FS+Planck than in BAO+Planck, as illustrated in  Fig.~\ref{fig:2D_gnu-sumnu} for $m_{\phi}=10\, \rm{eV}$.  
The left panel displays the two-dimensional constraints in the  $\sum m_{\nu}$--$g_{\nu}$  plane, showing that BAO data lead to tighter bounds on the sum of neutrino masses, mainly due to the breaking of degeneracies with $H_0$. The right panel presents the marginalized distributions of key cosmological parameters, showing that FS+Planck yields results consistent with those obtained from BAO+Planck. A similar trend is observed across the other mediator masses, confirming that, although the use of EdS kernels is not a fully valid approximation for some mediator masses, the dominant constraining power in the FS+Planck combination arises from the CMB data.

In addition, Fig.~\ref{fig:triangulars} confirms that the coupling parameter $g_\nu$ does not exhibit significant degeneracies with standard neutrino-sector parameters such as $\sum m_{\nu}$ and $N_{\rm eff}$, or with other cosmological parameters, indicating that the RNSI model introduces no additional projection effects than those present in the $\Lambda$CDM model. Moreover, the RNSI model does not seem to alleviate the preference for unphysical negative neutrino masses.


Overall, our findings highlight the capability of FS data, both alone and in combination with CMB observations, to constrain neutrino interactions in the resonant regime across a wide range of mediator masses, while maintaining compatibility with standard cosmological parameters.



\begin{table*}[htb]
\centering
\begingroup
\renewcommand{\arraystretch}{1.3}
\begin{tabular}{lllllll}
\toprule
\hline
\hline
Model &                               0.01\,eV &                                0.1\,eV &                                  1\,eV &               10\,eV &                                         100\,eV &            ${\rm \Lambda}$CDM \\
\hline
\textbf{FS+BBN} \\
$g_{\nu} \times 10^{14}$ &  --- &  --- &  ---  &  $<733$ &  --- &          --- \\
$\sum m_{\nu}\, [{\rm eV}]$ &                              --- &   --- &   $< 1.14$ &$< 1.61$ & $< 1.69$ &       $< 1.42$ \\
$N_{\rm eff}$ &                         --- &                         --- &                         $3.02\pm 0.27$ &       $2.98\pm 0.29$ &                                  $3.01\pm 0.27$ &  $2.99\pm 0.27$ \\
$\Omega_{m}$& --- & --- & $0.359^{+0.019}_{-0.021}   $& $0.355^{+0.020}_{-0.024}  $ &$0.350^{+0.019}_{-0.024}$ & $0.352^{+0.019}_{-0.023}   $\\
$\sigma_8$ & --- & --- & $0.675\pm 0.040 $ &$0.643^{+0.035}_{-0.040}   $ &$0.697^{+0.041}_{-0.046} $ & $0.699^{+0.041}_{-0.046}$\\
$H_0 [\kmsMpc]$     & --- &   --- &   $69.9\pm 1.9$ &  $69.0\pm 2.0$ &                                   $69.2\pm 2.0$ &   $69.1\pm 1.9$ \\
\hline

\textbf{FS+Planck} \\
$g_{\nu} \times 10^{14}$ & $<892$ & $<87.8$ & $<10.8$&$<32.4$& $<139 $&  --- \\
$\sum m_{\nu} \, [{\rm eV}]$                &  $<0.155 $ & $<0.172$& $<0.181$&$<0.177$& $<0.170$& $< 0.168$\\
$N_{\rm eff}$            & $3.16\pm 0.22$&$3.13\pm 0.23$&$2.97^{+0.17}_{-0.21}$&$2.97\pm 0.20 $&$3.00\pm 0.20$& $2.95\pm 0.22 $ \\
$\Omega_{m}$ &$0.314^{+0.008}_{-0.010}$&$0.314\pm 0.008$ & $0.316\pm 0.008$ &$0.317\pm 0.008$ & $0.315\pm 0.008$ & $0.316\pm 0.008$\\
$\sigma_8$& $0.815\pm 0.015$&$0.813\pm 0.014$& $0.803^{+0.014}_{-0.011}   $&$0.805^{+0.013}_{-0.012}$&$0.805^{+0.014}_{-0.012}$ & $0.804^{+0.014}_{-0.012}   $\\
$100\theta_s$&$1.0419\pm 0.0006$ & $1.0419\pm 0.0006 $&$1.0422\pm 0.0006$&$1.0430^{+0.0008}_{-0.0009}$& $1.0421\pm 0.0006 $&$1.0421\pm 0.0006$\\


\hline
\hline
\end{tabular}
\endgroup
\caption{
    Constraints on cosmological parameters for different combinations of datasets and mediator masses. On top are the CMB-independent constraints from the combination of FS and BBN, while the bottom shows the FS+Planck results. The last column corresponds to the base $\Lambda$CDM model with $\sumnu$ and $N_{\rm eff}$ as free parameters. Uncertainties are given at 68\%, while upper bounds are reported at 95\% confidence.
}
\label{tab:constraints}
\end{table*}

\section{\label{sec:conclusions} Conclusions}

In this work, we employed the effective field theory of large-scale structure to investigate, for the first time, resonant neutrino self-interactions by computing the redshift-space power spectrum multipoles up to mildly nonlinear scales. In combination with BBN, we derived CMB-independent constraints on the interaction coupling $g_\nu$. Additionally, we explored the impact of combining full-shape data with Planck.
Among the explored mediator masses, \( m_\phi = 10\,\mathrm{eV} \) and \( m_\phi = 1\,\mathrm{eV} \) draw particular attention due to their distinct phenomenological impact. For \( m_\phi = 10\,\mathrm{eV} \), FS+BBN data suffice to constrain the interaction, yielding \( g_\nu <7.33 \times 10^{-12} \) (95\% c.l.), which tightens to \( g_\nu < 1.75 \times 10^{-13} \) (95\% c.l.) when we combine FS with Planck data, representing more than an order of magnitude gain. The latter bound is also comparable to that obtained from the BAO+Planck combination in Ref. \cite{Venzor:2023aka} (\( g_\nu < 3.67 \times 10^{-13} \) at 95\% c.l.), further confirming the effectiveness of full-shape data in this regime and mediator mass.
On the other side, at \( m_\phi = 1\,\mathrm{eV} \), where Planck data are particularly sensitive to the interaction, FS+Planck yields the strongest overall constraint: \( g_\nu < 1.08 \times 10^{-13} \) (95\% c.l.), tightening by around 28\% our previous constraint of $g_\nu < 1.50 \times 10^{-13}$ (95\% c.l.), obtained for that mediator mass from the combination of BAO and Planck \cite{Venzor:2023aka}. 
These two cases illustrate the complementarity between datasets in testing neutrino interactions and highlight the power of full-shape information beyond what is captured by BAO alone, especially when combined with CMB observations.

In summary, this work places the strongest cosmological constraints to date on resonant neutrino self-interactions across a wide range of mediator masses, using full-shape galaxy power spectrum data.
Importantly, we show in Fig. \ref{fig:triangulars} that the RNSI model does not exhibit significant degeneracies with $N_{\rm eff}$ and $\sumnu$, or other cosmological parameters. 
This allows for independent and robust bounds of the coupling parameter.

Resonant neutrino self-interactions provide an alternative framework to explore new physics in the neutrino sector, particularly through their impact on astrophysical and cosmological observables. Previous studies have focused primarily on models with heavy mediators, which often lead to bimodal posteriors for the interaction strength, especially when analyzed with CMB--Planck data. 
However, more recent results have shown that this bimodality can be suppressed when large-scale structure data are considered \cite{he2023self,Camarena:2024daj,He:2025jwp,Poudou:2025qcx,Camarena2023}.
In contrast, the RNSI model for the explored masses does not exhibit such bimodality, even when using Planck data alone, and throughout this paper we prove that including FS data is consistent with previous results of this interaction regime. 
For instance, we have placed new cosmological bounds on the model parameters by conducting a Bayesian analysis for two different dataset combinations: FS+BBN and FS+Planck, enabling a broad comparison of the impact of low and early redshift data on the parameter space.

In this work, we have used the Planck \texttt{Plik} 2018 likelihood \cite{Planck:2018nkj}, however Ref. \cite{Poudou:2025qcx} showed that newer likelihoods of the Planck data change the constraints in the self-interacting parameters, at least for the heavy mediator case. CamSpec \cite{Rosenberg:2022sdy} reduces the constraints in $\log_{10}G_{\rm eff,\nu}$ to a 6\% factor for the heavy mediator. We expect similar changes in our constraints, that would be made in future work together with a DESI full-shape analysis \cite{DESI:2024jis} once these data become fully public.

It is noteworthy that some UV complete models constructed to generate large neutrino self-interactions via heavy mediators \cite{Berbig2020,He_2020,Lyu2021} suggest mediator masses consistent with those investigated herein. 
The resonant behavior of these interactions during the early universe makes cosmological observables an optimal source of knowledge for this kind of new physics.
Nevertheless, consistent with prior discussion, our RNSI analysis yields results that are not strongly indicative of new physics beyond the Standard Model.

\acknowledgments

The authors thankfully acknowledge Cl\'uster de Superc\'omputo Xiuhcoatl (Cinvestav) for the allocation of computer resources.
HN acknowledges support by SECIHTI grant CBF2023-2024-162 and PAPIIT IA101825.
GG-A and JV acknowledge support from SECIHTI postdoctoral fellowships.

\bibliographystyle{JHEP} 

\bibliography{references}

\end{document}